\documentclass[runningheads]{llncs}
\usepackage{graphicx}
\usepackage{xcolor}

\begin{document}
\title{Contour Completion by Transformers and Its Application to Vector Font Data\thanks{This research was partially supported by MEXT-Japan (Grant No. JP22H00540 and JP23K16949).}}
%
\titlerunning{Contour Completion by Transformers}
%
\author{Yusuke Nagata\inst{1}\orcidID{0009-0001-0498-4604} \and
Brian Kenji Iwana\inst{1}\orcidID{0000-0002-5146-6818} \and
Seiichi Uchida\inst{1}\orcidID{0000-0001-8592-7566}}

\authorrunning{Yusuke Nagata et al.}

\institute{Kyushu University, Fukuoka, Japan \\
\email{yusuke.nagata@human.ait.kyushu-u.ac.jp}\\
\email{\{iwana,uchida\}@ait.kyushu-u.ac.jp}}
\maketitle              
\begin{abstract}
In documents and graphics, contours are a popular format to describe specific shapes. For example, in the True Type Font (TTF) file format, contours describe vector outlines of typeface shapes. Each contour is often defined as a sequence of points. In this paper, we tackle the contour completion task. In this task, the input is a contour sequence with missing points, and the output is a generated completed contour. This task is more difficult than image completion because, for images, the missing pixels are indicated. Since there is no such indication in the contour completion task, we must solve the problem of missing part detection and completion simultaneously. We propose a Transformer-based method to solve this problem and show the results of the typeface contour completion.

\keywords{Vector font generation \and Contour completion \and Transformer.}
\end{abstract}
\section{Introduction}
\label{sec:intro}

Contours are a popular method of representing vector objects in document and graphics recognition. 
For example, digital-born fonts are often formatted using the True Type Font (TTF) file format. 
This format consists of a sequence of contour points for each character in the font set. 
The contour points, called \textit{control points}, contain a point identifier, contour identifier, two-dimensional coordinates, and an on-curve/off-curve flag. The point identifier identifies the location in the sequence, the contour identifier identifies the particular contour that the point lies on, the coordinates describe the coordinates on a relative scale and the flags determine whether the point is on a curve or an angle.

\begin{figure}[t]
\centering
\begin{minipage}[b]{0.3\textwidth}
    
    \includegraphics[width=1.0\linewidth]{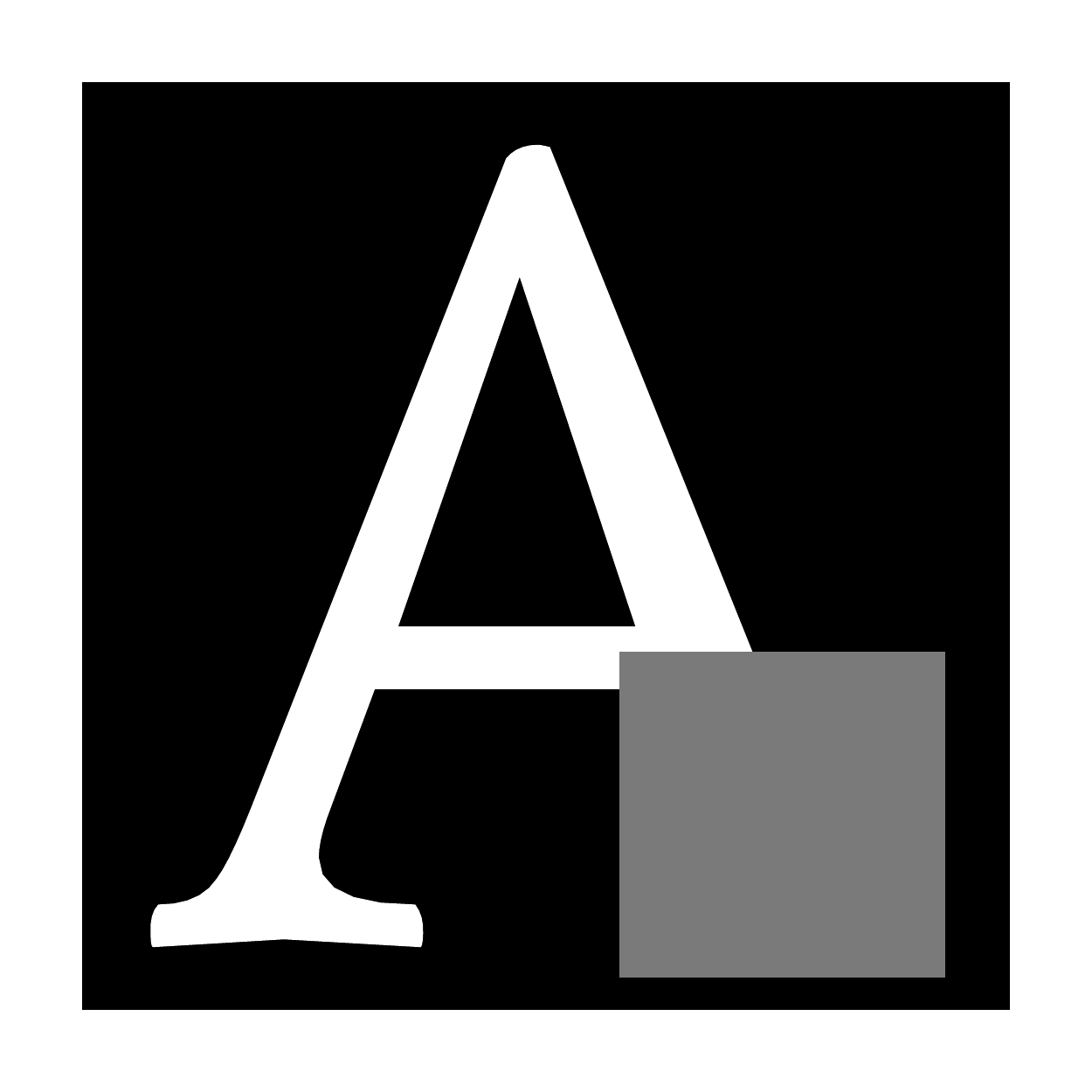}\\
    (a) Image-based completion problem
\end{minipage}
\hspace{2cm}
\begin{minipage}[b]{0.3\textwidth}
    \centering
    \includegraphics[width=1.0\linewidth]{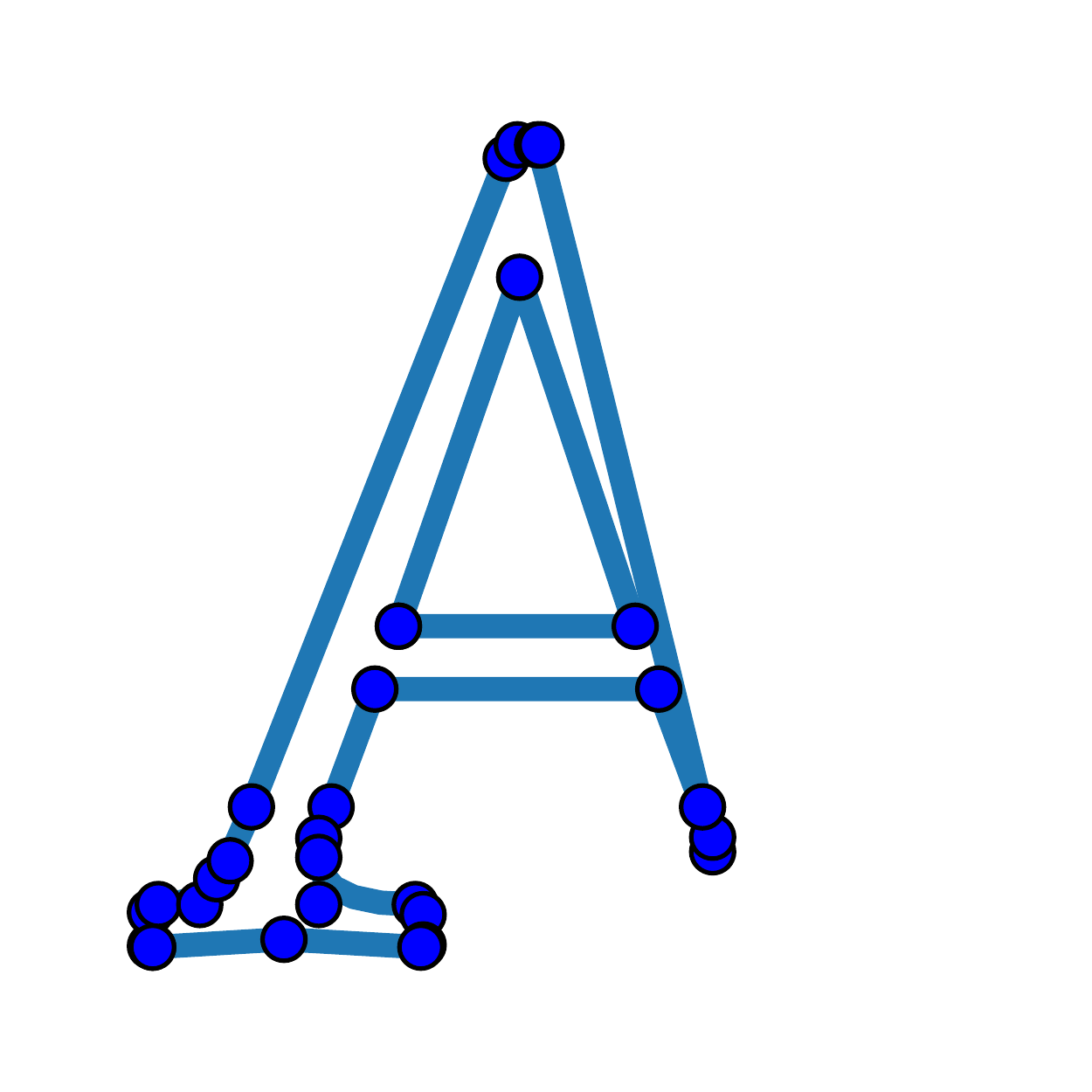}\\
    (b) Indicator-free contour completion problem
\end{minipage}
\caption{Comparison between image-based completion and vector-based completion. Image-based completion uses a bounding box or mask to indicate the missing pixels. The proposed task does not have any indication of where the missing control points are.}
\label{fig:completion}
\end{figure}

In this paper, we perform vector completion of vector character contours using TTF data. 
However, unlike other completion problems, in our problem set, there is no indication of which parts of the contour data are missing.
For example, as shown in Fig.~\ref{fig:completion}~(a), in standard image completion problems, the missing pixels are specified with a mask or fixed pixel values (e.g., gray or negative values). 
For time series, interpolation and regression are similar; they require knowledge of the location of predicted data.
Conversely, we perform indication-free contour completion in that there is no indication, no mask, and no information about the missing coordinates. 
Therefore, the problem set is more difficult to solve due to needing to identify the missing points and estimate the missing values simultaneously.

There are two reasons for addressing the character contour completion problem.
The first is the application value.
Consider an application that converts bitmap character information into vector contour information (i.e., raster-vector transformation). 
In this application, if the character is partially overlapped, hidden by a design element, damaged, or incomplete, the result will be poor. 
By using automatic contour completion, such cases might be resolved. 
Also, in font design, automatic completion is possible without all of the contour points.
The second is the suitability for machine learning. 
There are a vast number of character fonts, and TTF is the most common format.
It is possible to leverage a large amount of data to train robust models to learn the trends in font and character tendency.

We consider the missing contour completion problem as a problem of converting a missing sequence into a completed sequence. 
To solve this problem, we propose using an Encoder-Decoder Transformer network~\cite{vaswani2017attention}.
Transformers were originally proposed for the Natural Language Processing (NLP) domain. 
However, they have been expanded to use for images~\cite{dosovitskiy2020image} and sequences~\cite{wen2022transformers}. 
The Transformer is trained to convert an input character contour sequence with missing control points (unknown to the Transformer) into a complete contour sequence.

Transformers have two major advantages for this problem. 
The first is the ease of handling series of varying lengths. 
Being able to tackle series of varying lengths is important because the contour sequences are not fixed in length. 
In our model, neither the input nor the output could have a fixed length. 
The second advantage is due to self-attention being suitable for our application. 
Character contours have structures that are both global and local to specific areas. 
For example, serifs exist in specific parts of characters. 
If the character is missing a serif, then the contour completion model needs to recognize that the serif is missing and reconstruct the serif based on the other serifs in the character. 
Thus, because Transformers have global receptive fields, we believe they are suitable for character contour completion. 
In fact, the font style recognition task has shown that high accuracy can be achieved by using contour data with  Transformers~\cite{nagata2022}.

To evaluate the performance of the proposed method, we conducted a contour completion experiment using fonts collected from Google Fonts~\cite{googlefonts}.
We formulated two problem setups: random deletion and burst deletion. 
In random deletion, randomly selected contour points were deleted. 
This is evaluated to demonstrate that the proposed method can identify and repair individual points. 
In burst deletion, large segments of the contour sequences are removed. 
Burst deletion provides a difficult problem because there is less information for the model to infer the correct contour from.

The main contributions of this paper are as follows.
\begin{itemize}
\item We use a Transformer to simultaneously identify missing control points in character contours and estimate the missing values. 
\item We propose the use of a multifaceted loss that considers the different features in the TTF file format for vector fonts.
\item We trained the above model using character data collected from Google Fonts and quantitatively and qualitatively evaluated the extent to which contour completion is possible.
\end{itemize}

\section{Related work}
\label{sec:related}
\subsection{Transformers}

Transformers~\cite{vaswani2017attention} are neural networks that use layers of Multi-Head Self-Attention and fully-connected layers and are designed for sequences recognition.
They were originally proposed for Natural Language Processing (NLP)~\cite{vaswani2017attention} such as translation and text classification.
They have had many successes in the NLP field~\cite{kenton2019bert,liu2019roberta}.

However, in recent years, Transformers have been applied to other fields such as image recognition~\cite{khan2021survey} and time series recognition~\cite{wen2022transformers}. 
To do this, instead of word token sequences for NLP, other embedding methods are used to create the input sequences. 
For example, Vision Transformers (ViT)~\cite{dosovitskiy2020image} use sequences of patched embedded in vector space.
For time series, many works use a linear embedding of time series elements along with positional embeddings for the inputs to the Transformers~\cite{li2019enhancing,zhou2021informer}.

\subsection{Vector Font Generation} \label{sec:related_vec}

There have been attempts related to font generation using vector formats. 
Campbell et al.~\cite{Campbell2014} pioneered outline-based font generation using a fixed-dimensional font contour model as a low-dimensional manifold. 
Lopes et al.~\cite{lopes2019learned} proposed a Variational Autoencoder (VAE) based model that extracts font style features from images and uses a Scalable Vector Graphic (SVG) based decoder to convert it to SVG format. 
In a similar work, Reddy et al.~\cite{reddy2021im2vec} proposed Im2vec which generates a vector from an image.  
DeepSVG~\cite{carlier2020deepsvg} is a font generation model using a hierarchical generative network to create SVG format fonts. 
DeepVecFont~\cite{wang2021deepvecfont} is a network that supports both image and sequential fonts using an encoder-decoder structure. 
In another work by Reddy et al. ~\cite{reddy2021multi}, they choose not to use SVG and instead learn font representations defined by implicit functions.

Unlike these font generation methods, we are focusing on automatic font repair and completion. 
Specifically, instead of generating vector fonts from images, we use sequence-to-sequence font generation to restore missing elements of fonts.
In addition, vector-based font generation methods typically have indications of where the missing data is or how much is missing. 
To the best of our knowledge, none of the previous methods simultaneously detect missing points and fill in the values.


\section{Missing Contour Completion with  Transformers}

\begin{figure}[t]
\centering
   \includegraphics[width=.95\linewidth]{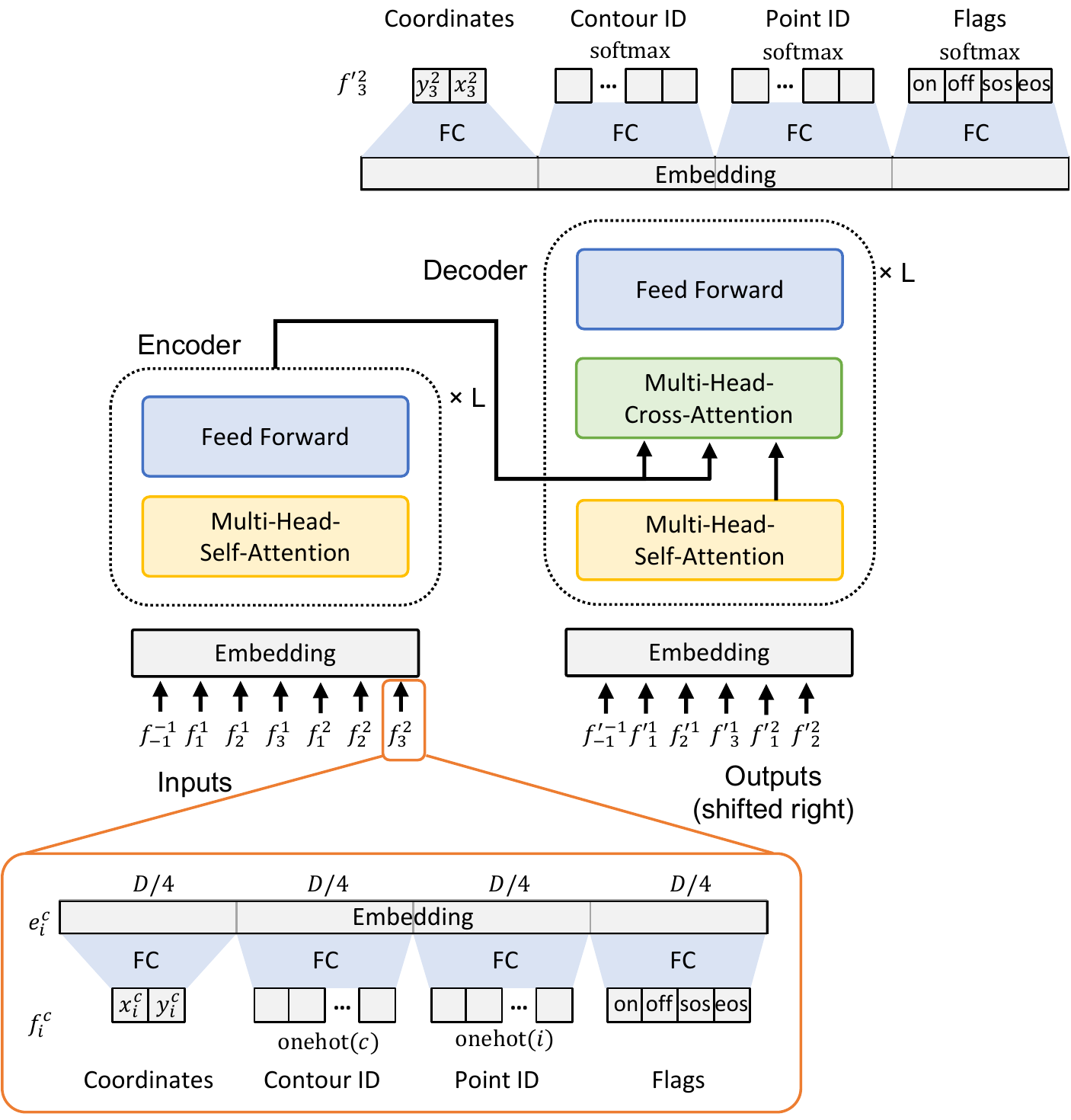}\\
\caption{Overview of the proposed Encoder-Decoder Transformer for contour completion.}
\label{fig:overview}
\end{figure}

We propose using a Transformer to automatically complete contour data when given a contour with missing points. 
Fig.~\ref{fig:overview} shows an overview of the proposed method.

\subsection{Contour Data Representation}
\label{sec:outline_data}

To represent the contours of fonts as a sequence for the Transformer, we use the True Type Font (TTF) file format of each font. 
The TTF files contain the vector information of fonts in the form of the contours of each character in that font. 
Specifically, the outlines of each character are expressed as a variable number of point coordinates. 
Each point is represented by a 5-dimensional vector $f^c_i$, which consists of $x$ and $y$ coordinates, a Contour ID, a Point ID, and a curve flag.
The $x$ and $y$ coordinates are the relative locations of the  control point. 
The Contour ID $c\in\{-1, 1,\ldots,C\}$ and Point ID $i\in\{-1, 1,\ldots,I^c\}$ are identifiers that label the contour number $C$ and point number $I^c$, respectively. 
The IDs, $c=-1$ and $i=-1$, are used as placeholders for the start-of-sequence and end-of-sequence tokens.
In the contour data, $f^c_1=f^c_{I^C}$, since the start and end points of each contour indicate exactly the same position. Thus, the character data is represented by a set of $N=\sum_c P^c$ vectors, which are input to the Transformer in the order $\mathbf{f}=f^{-1}_{-1},f^1_1,\ldots,f^1_{I^1},f^2_1,\ldots,f^c_i,\ldots,f^C_{I^C}$.

Finally, the curve flag is an indicator if the control point is \textit{on-curve} or \textit{off-curve}. 
If the control point is on-curve, then B-spline is used to define the curvature. 
If it is off-curve, then an angle is used. 
Furthermore, for the purpose of our input data representation, the start-of-sequence and end-of-sequence flags are grouped with the curve flags.

\subsection{Encoder-Decoder Transformer for Contour Completion}
\label{sec:overview}

We propose the use of an Encoder-Decoder Transformer to perform contour completion, as shown in Fig.~\ref{fig:overview}. 
The input to the Transformer is a character contour with missing control points and the output is the completed character contour. 
This is analogous to a translation task in that a variable length input is translated to a variable length completed contour output. 
Notably, the input has no indication of the location of missing control points. 
Therefore, the proposed method simultaneously detects missing control points and infers the values of the missing control points. 

The proposed method is different from the original Transformer~\cite{vaswani2017attention} in three ways:
\begin{itemize}
    \item We adapt the token embedding of the Transformer for the use of vector contours, i.e. TTF file format.
    \item We consider the problem as a multitask problem to predict the coordinates, Contour ID, Point ID, and flags, separately.
    \item We propose a new loss to optimize the Transformer in order to consider each task.
\end{itemize}

\subsubsection{Embedding}

To adapt the standard token embedding for contours, each element of $\mathbf{f}$ is embedded into a $D$-dimensional vector. 
As shown in Fig.~\ref{fig:overview}, to do this, we split the features of the TTF file format into four groups, the coordinates, the Contour ID, the Point ID, and the flags. 
Each group is then used with a fully-connected (FC) layer to embed it into a $D/4$-dimensional vector. 
The resulting four $D/4$-dimensional vectors are concatenated to create a $D$-dimensional vector embedding $e_i^c$ for each input $f_i^c$ in $\mathbf{f}$. 
The sequence of embeddings is the input to the Transformer.

Namely, the 2-dimensional coordinate values $(x^c_i, y^c_i)$ are embedded using the FC layer to the first segment of $e_i^c$. 
Next, the Contour ID and Point ID are converted to one-hot representations, and then embedded using the FC layer.
The final $D/4$ dimensions use a 4-dimensional one-hot input representing an on-curve flag, an off-curve flag, a start-of-sequence (sos) flag, and an end-of-sequence (eos) flag.

\subsubsection{Encoder-Decoder Structure}

The Transformer is arranged in an Encoder-Decoder structure, similar to how it is used in NLP. 
The model is divided into an Encoder and a Decoder.
The input of the Encoder is the previously mentioned sequence of embeddings. 
Similar to a standard Transformer, the Encoder is made of $L$ Transformer layers that each contain a Multi-Head Self-Attention layer and a Feed Forward layer.
The output of the Encoder is an element-wise embedding that represents each input in a sequence of $N$ number of $D$-dimensional vectors.

The Decoder is structured similarly to Vaswani et al.~\cite{vaswani2017attention}.
The input of the Decoder is the generated sequence of control points, shifted right. 
The output of the Decoder is the next predicted control point in the sequence. 
To connect the Encoder and the Decoder, the output of the Encoder is used as the key $K$ and value $V$ of the Multi-Head Cross-Attention layer. 
The query $Q$ is the out of the Multi-Head Self-Attention layer of the Decoder.
Much like the Encoder, the Decoder also has $L$ number of Transformer layers. 

\subsubsection{Output}

One proposed modification of the Transformer is to use a multitask prediction for the output of the Decoder.
Specifically, as shown in the top of Fig.~\ref{fig:overview}, the $D$-dimensional vector representation of the Decoder is divided into four parts, each part relating to the coordinates, Contour ID, Point ID, and flags, respectively. 
Then, an FC layer is used with each part to determine the prediction of each part. 
The FC layer for the coordinates has two nodes, one for $x^c_i$ and one for $y^c_i$. 
The FC layers for the Contour ID, Point ID, and flags use softmax to perform the one-hot prediction.

\subsubsection{Loss Functions}
\label{sec:loss}

In order to train the network, we propose a multifaceted loss to consider all of the information in the TTF file format.
The loss function consists of four terms: contour loss $\mathcal{L}_\mathrm{contour}$, point loss $\mathcal{L}_\mathrm{point}$, coordinate loss $\mathcal{L}_\mathrm{coord}$, and flag loss $\mathcal{L}_\mathrm{flag}$. 
The contour loss $\mathcal{L}_\mathrm{contour}$ ensures that the predicted point is part of the correct contour. 
It is defined as the cross entropy loss between the predicted contour ID $q_c$ and the true contour ID $p_c$, or:
\begin{equation}
\mathcal{L}_\mathrm{contour} = -\frac{1}{C}\sum_{c=1}^C p_{c} \log(q_{c}),
\end{equation}
where $C$ is the number of contours. 
The point loss $\mathcal{L}_\mathrm{point}$ is similar to the contour except that is for the control point ID, or:
\begin{equation}
\mathcal{L}_\mathrm{point} = -\frac{1}{I}\sum_{i=1}^{I^c} p_{i} \log(q_{i}),
\end{equation}
where $I^c$ is the number of control points in the contour $c$, $p_i$ is the true point ID, and $q_i$ is the predicted point ID. 
This loss makes sure that the model learns the location of the missing control point in the sequence.
The coordinate loss is the difference between the predicted coordinates $\hat{y_i}$ and the true coordinates $y_i$. For this loss, we use L1 loss, or:
\begin{equation}
\mathcal{L}_\mathrm{coord} = \sum_{i=1}^{I^c}{|y_i-\hat{y_i}|}.
\end{equation}
Finally, the flag loss $\mathcal{L}_\mathrm{flag}$ is the loss between the four flags $p_u$ and the predicted four flags $q_u$, or:
\begin{equation}
    \mathcal{L}_\mathrm{flag} = -\frac{1}{4}\sum_{u=1}^4 p_{u} \log(q_{u}).
\end{equation} 
The four losses are summed in the total loss $\mathcal{L}$:
\begin{equation}
\mathcal{L}=\mathcal{L}_\mathrm{contour}+\mathcal{L}_\mathrm{point}+\mathcal{L}_\mathrm{coord}+\mathcal{L}_\mathrm{flag}.
\end{equation}
The Transformer is trained using the total loss $\mathcal{L}$.

\section{Experiment}
\label{sec:experiment}
\subsection{Dataset}
\label{sec:dataset}

\subsubsection{Font Database}

For the experiments, we used the Google Fonts~\cite{googlefonts} database because it is one of the most popular font sets for font analysis research~\cite{roy2020stefann,srivatsan2019deep,srivatsan2021scalable}.
It also contains a large number of fonts in TTF file format.
They are annotated with five font style categories (\textit{Serif}, \textit{Sans-Serif}, \textit{Display}, \textit{Handwriting}, and \textit{Monospace}). 
From Google Fonts, we selected the Latin fonts that were used in STEFANN~\cite{roy2020stefann}. We then omitted \textit{Monospace} fonts because they are mainly characterized by their kerning rule (i.e., space between adjacent letters); therefore, some \textit{Monospace} fonts have the same style of another font type, such as \textit{Sans-Serif}.
We set the upper limit of Contour IDs and Point ID to $C=4$ and $P=102$, respectively.
Consequently, we used 489 \textit{Serif} fonts, 1,275 \textit{Sans-Serif}, 327 \textit{Display}, and 91 \textit{Handwriting} fonts. The fonts are split into three font-disjoint subsets; that is, the training set with 1,777 fonts, the validation set with 200 fonts, and the test set with 205 fonts.

\subsubsection{Simulation of Missing Data}

\begin{figure}[t]
\begin{minipage}[b]{0.3\textwidth}
    \centering
    \includegraphics[width=1.0\linewidth]{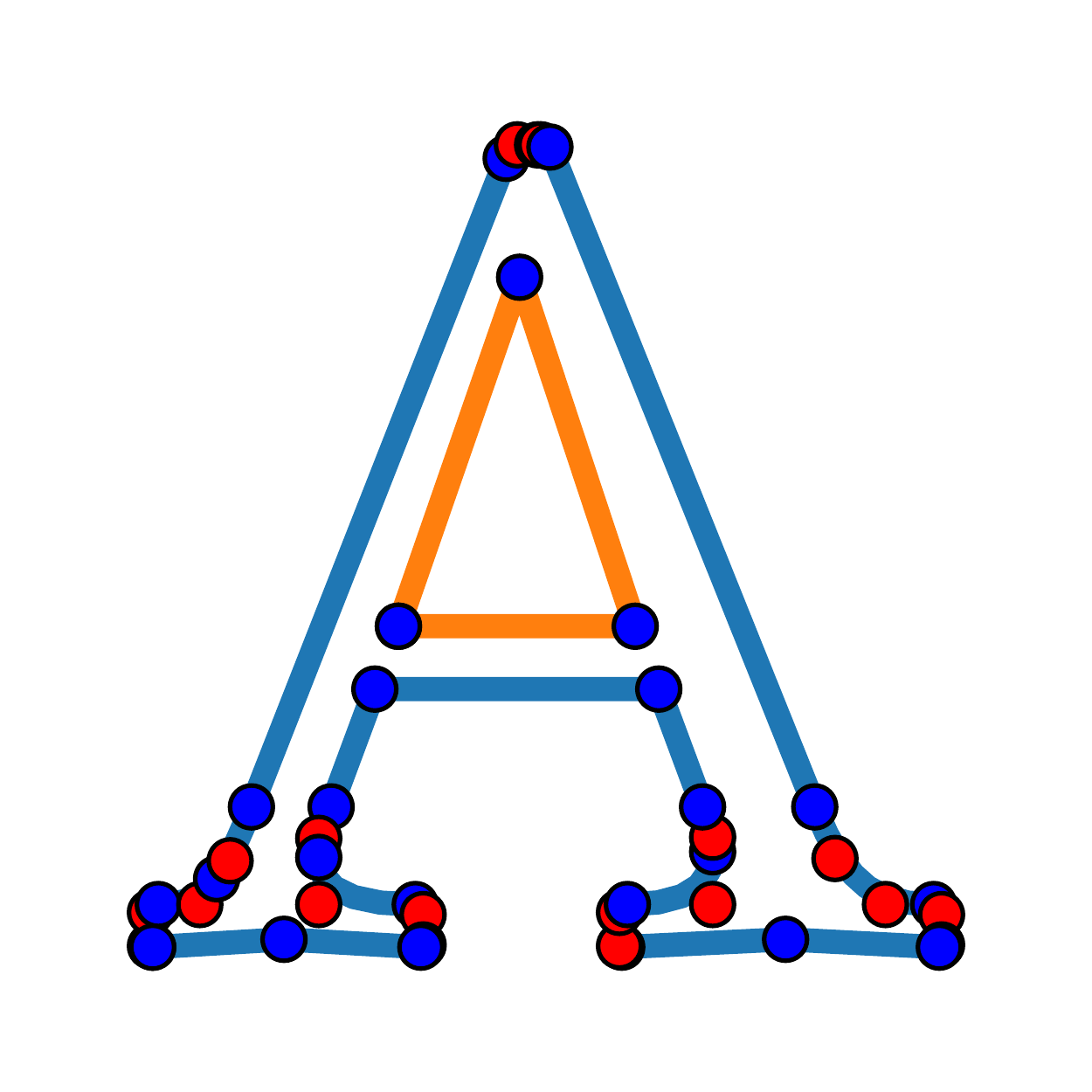}\\
    (a) Completed contour
\end{minipage}
\hfill
\begin{minipage}[b]{0.3\textwidth}
    \centering
    \includegraphics[width=1.0\linewidth]{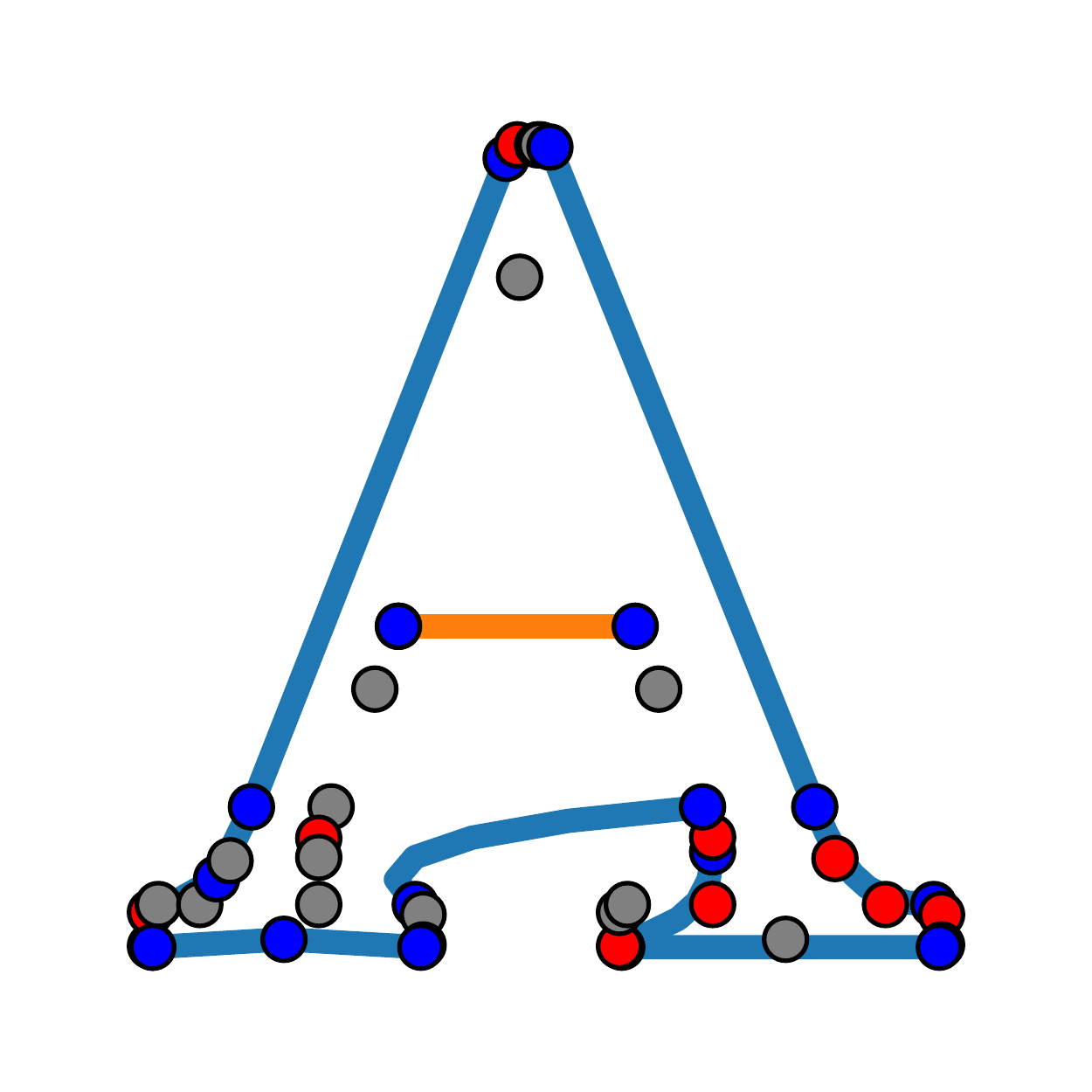}\\
    (b) Random deletion
\end{minipage}
\hfill
\begin{minipage}[b]{0.3\textwidth}
    \centering
    \includegraphics[width=1.0\linewidth]{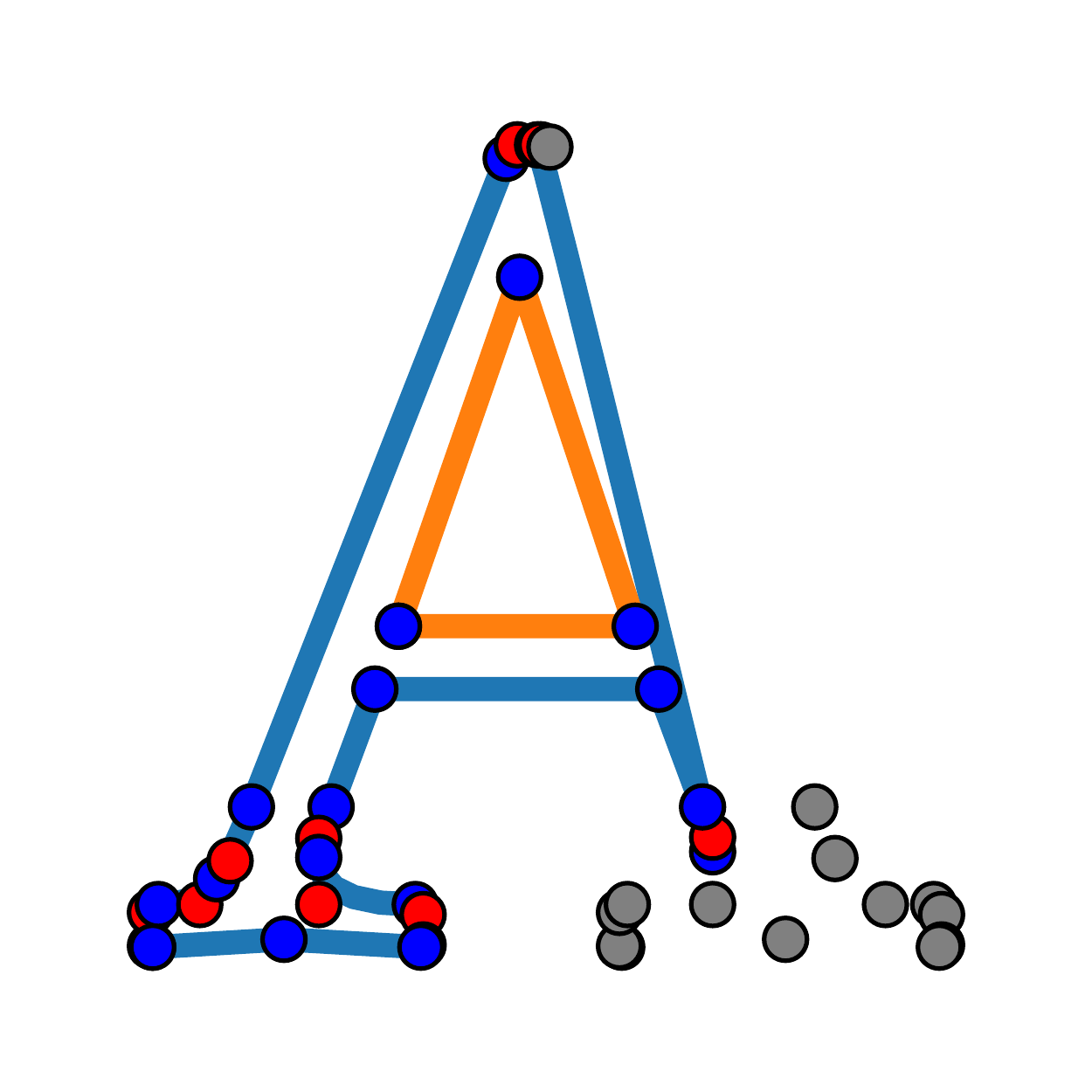}\\
    (c) Burst deletion
\end{minipage}
\caption{Comparison of the deletion methods used to simulate missing data. The red points are on-curve points, the blue points are off-curve points, and the gray points are deleted control points.}
\label{fig:deletion}
\end{figure}

To simulate missing contour data, the control points were removed from the font contours. 
We modify the fonts in two ways, \textit{random deletion} and \textit{burst deletion}. 
In random deletion, as shown in Fig.~\ref{fig:deletion}~(b), $D$ points are randomly removed from the characters. 
For burst deletion, intervals or segments of control points are removed.
To perform burst deletion, as shown in Fig.~\ref{fig:deletion}~(c), a random control point is removed along with the $D-1$ points surrounding it. 
In the experiments, we compare different deletion rates. 
Thus, the number of deleted points $D$ is determined by $D=rN$, where $r$ is the deletion rate and $N$ is the total number of control points in a character. 

\subsection{Implementation Details}
\label{sec:model_detail}

The Transformer used in the experiment is shown in Fig.~\ref{fig:overview}. 
The Encoder and Decoder use $L=4$ number of the Transformer layers each. 
We set Mutli-head Attention heads as $M=6$ and The internal dimensionality of FC layers set at $D=256$.
To train the network, a batch size of 400 is used with an Adam~\cite{kingma2014adam} optimizer with an initial learning rate of 0.0001. 
Training of the network was stopped by no decrease in validation loss for 30 epochs.

\subsection{Comparative Method}
To evaluate the proposed method, we use a Transformer-Encoder model. 
This model is similar to the Encoder used in the proposed method, except the number of layers and Multi-Head Self-Attention layers are increased to be equal to the proposed method. 
The model receives the same embedding as the proposed method and the output is performed in the same manner as the proposed method. 
The difference between the two is that the comparative model is like a standard neural network in that the layers are stacked and not parallel like the Encoder-Decoder Transformer. 
For fairness, the network has the same hyperparameters and is trained with the same optimizer and training scheme as the proposed method.

Note that, unlike the proposed method which can have different length inputs and outputs, the comparative method can only predict same length outputs as the inputs.
This means that a placeholder must be used for the missing control points. 
This is much like traditional completion tasks where the missing information is explicitly indicated. 
Thus, only the value prediction is required, which reduces the difficulty of the problem.

\subsection{Quantitative Evaluation }

\begin{figure}[t]
\begin{minipage}[b]{0.49\textwidth}
    \centering
    \includegraphics[width=1.0\linewidth]{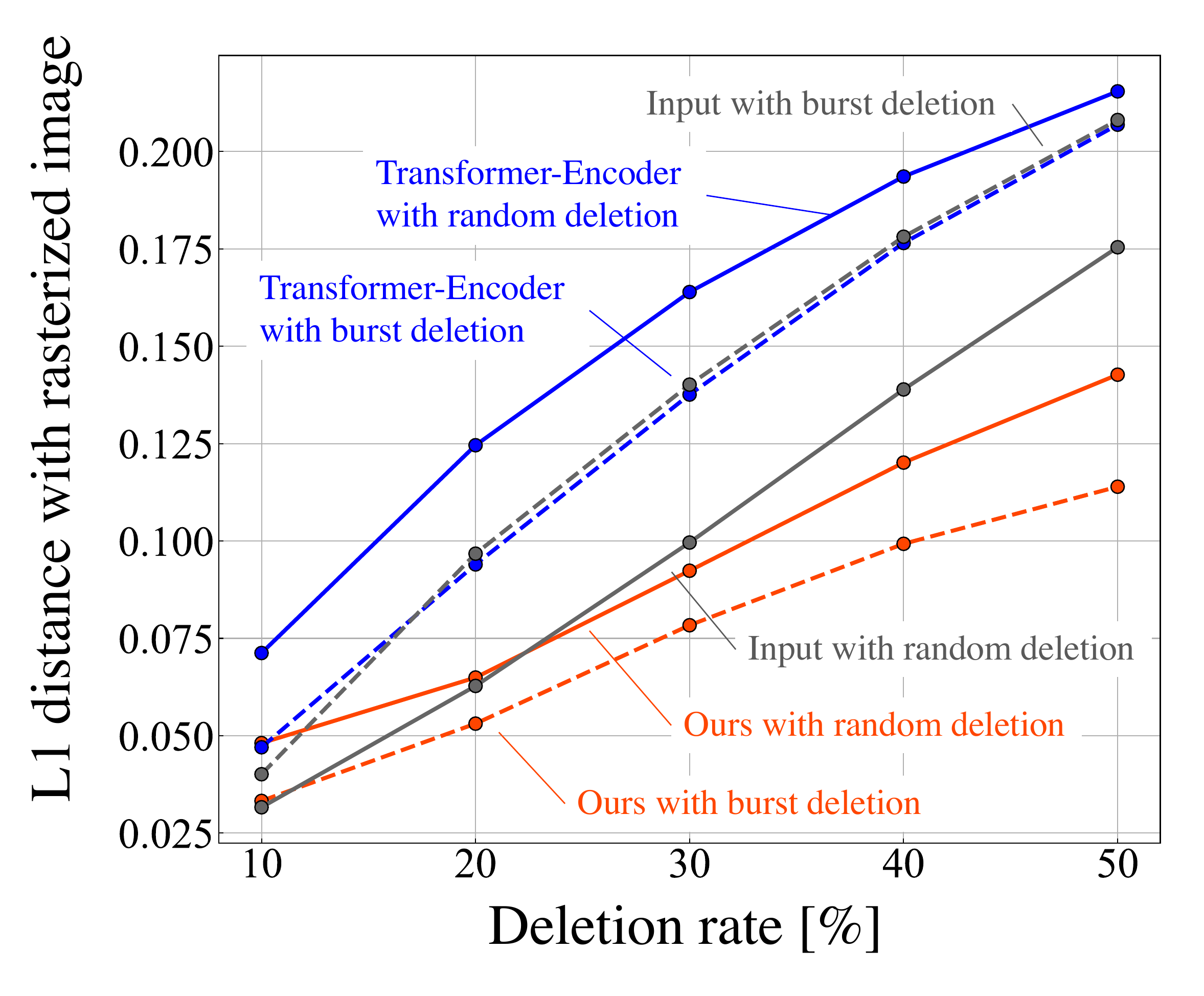}\\
    (a) L1 distance between ours and the comparative method
\end{minipage}
 \hspace{0.05\linewidth}
\begin{minipage}[b]{0.49\textwidth}
    \centering
    \includegraphics[width=1.0\linewidth]{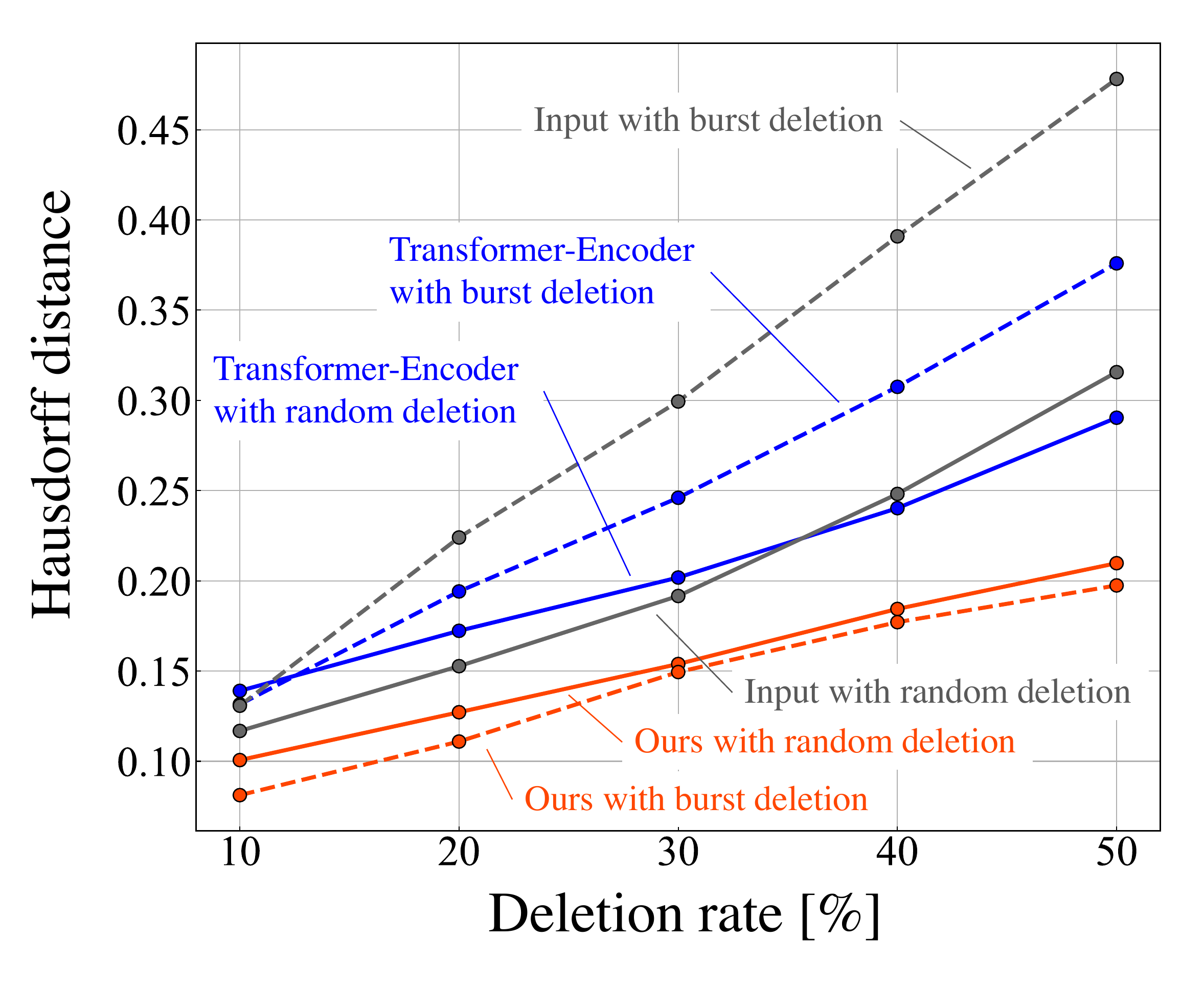}\\
    (b) Hausdorff distance between ours and the comparative method
\end{minipage}
\caption{L1 distance with rasterized image and Hausdorff distance}
\label{fig:gragh}
\end{figure}

To compare the proposed method with the comparative method, we evaluate them with different deletion rates with both random deletion and burst deletion. 
Fig.~\ref{fig:gragh} shows the L1 distance and Hausdorff distance~\cite{taha2015efficient} between the contour completion methods and the ground truth.
To calculate the L1 distance $d_\mathrm{L1}$, the contours are rasterized into bitmap images of $250\times250$ pixels. 
The bitmaps are then compared using:
\begin{equation}
  \mathrm{L1}(\mathbf{f},\mathbf{\hat{f}}) = \sum_{(x, y)}{|R({\mathbf{f}})_{(x,y)} - R({\mathbf{\hat{f}}})_{(x,y)}|},
\end{equation}
where $(x,y)$ is the coordinate of each pixel, $R(\mathbf{f})$ is the rasterized font, and $R({\mathbf{\hat{f}}})$ is the rasterized predicted font.
For the Hausdorff distance, the sequences are considered point clouds. 
The Hausdorff distance can be calculated as the distance between two point clouds with no point-to-point correspondence. 
It is widely used in 3D point cloud generation.
It is defined as:
\begin{equation}
    \check{H}(\mathbf{f},\mathbf{\hat{f}})=\max_{f\in{\mathbf{f}}}\{\min_{f'\in{\mathbf{\hat{f}}}}\{\|f, f'\|\}\},
\end{equation}
\begin{equation}
H(\mathbf{f},\mathbf{\hat{f}})=\max\{\check{H}(\mathbf{f},\mathbf{\hat{f}}),\check{H}(\mathbf{\hat{f}},\mathbf{f})\},
\end{equation}
where $\|\cdot\|$ is the L2 distance between the nearest points in the respective sequences, and $f$ is a point in the ground truth $\mathbf{f}$ and $f'$ is a point in the predicted $\mathbf{\hat{f}}$. 
For both distances, smaller distances mean that the contours are more accurately reproduced.

\begin{figure*}[!t]
\centering
\includegraphics[width=1.0\linewidth]{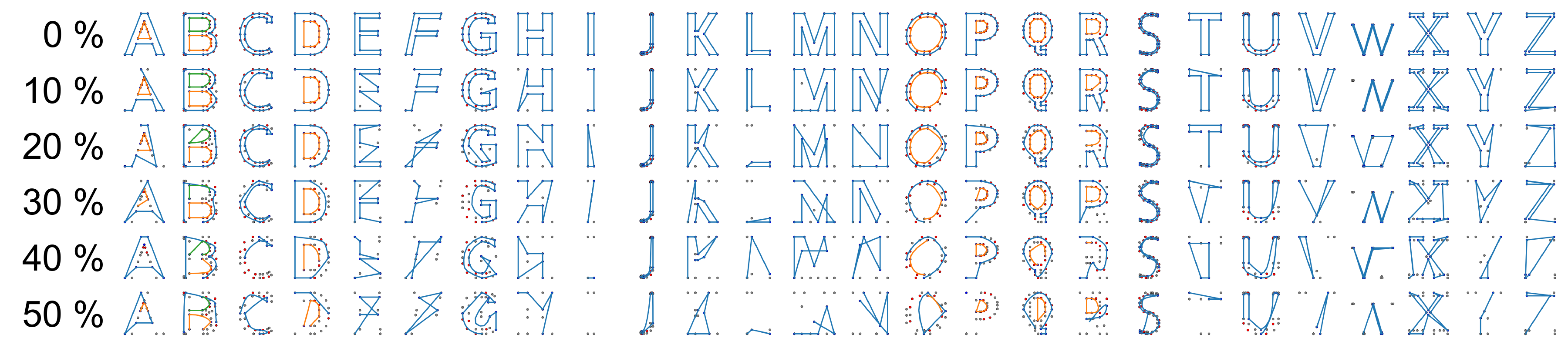}\vspace{-1mm}
(a)~{Input (Random deletion)}
\vspace{4mm}

\includegraphics[width=1.0\linewidth]{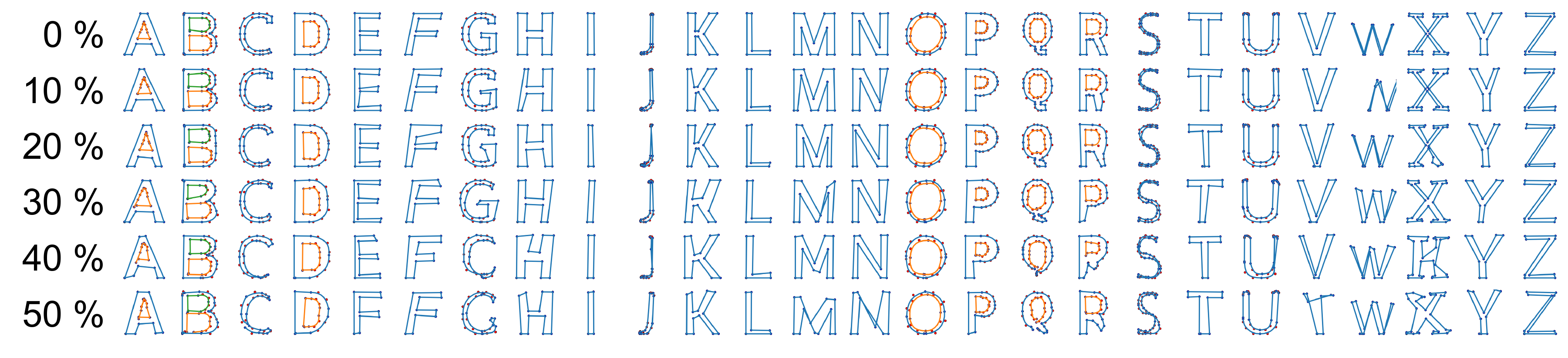}\vspace{-1mm}
(b)~{Proposed method}
\vspace{4mm}

\includegraphics[width=1.0\linewidth]{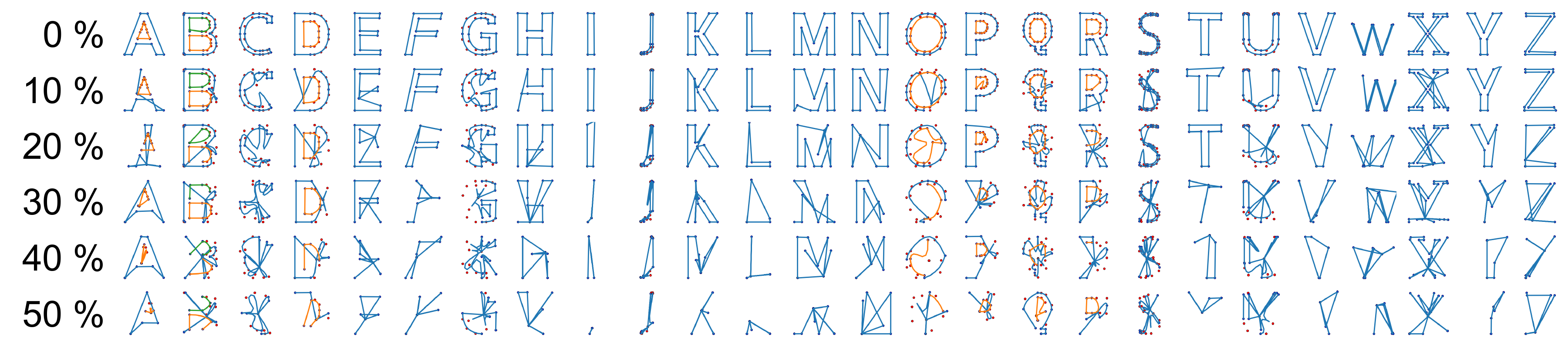}\vspace{-1mm}
(c)~{Comparative method}
\caption{The results on random deletion. The columns are characters in the same font style, and the rows are the same missing percentages. The first row is the ground truth complete contour and second to sixth row have a deletion ratio of 10\% to 50\%, respectively. The colors indicate the Contour ID.}
\label{fig:random_results}
\end{figure*}

For the L1 distance and the Hausdorff distance evaluations, the proposed method is better than the comparative method at all deletion rates and deletion methods. 
Despite burst deletion having a very high L1 distance and Hausdorff distance, the proposed method was able to perform the contour completion task very well. 
However, at 10\% and 20\%, the L1 distance for the proposed method is larger than the input. 
This is because, at low deletion rates, most of the control points are intact. 
That means the rasterized characters are mostly correct, whereas the generated contours might have very slight translations causing the pixel-based distance to be larger than the inputs with incomplete contours.

Also, another interest interesting result is that both Transformers were able to have better results for L1 distance on the input with burst deletion than random deletion despite the burst deletion having higher distances originally. 
This means that quantitatively, the Transformers were able to recover large continuous segments better than randomly deleted points.

\begin{figure*}[!t]
\centering
\includegraphics[width=1.0\linewidth]{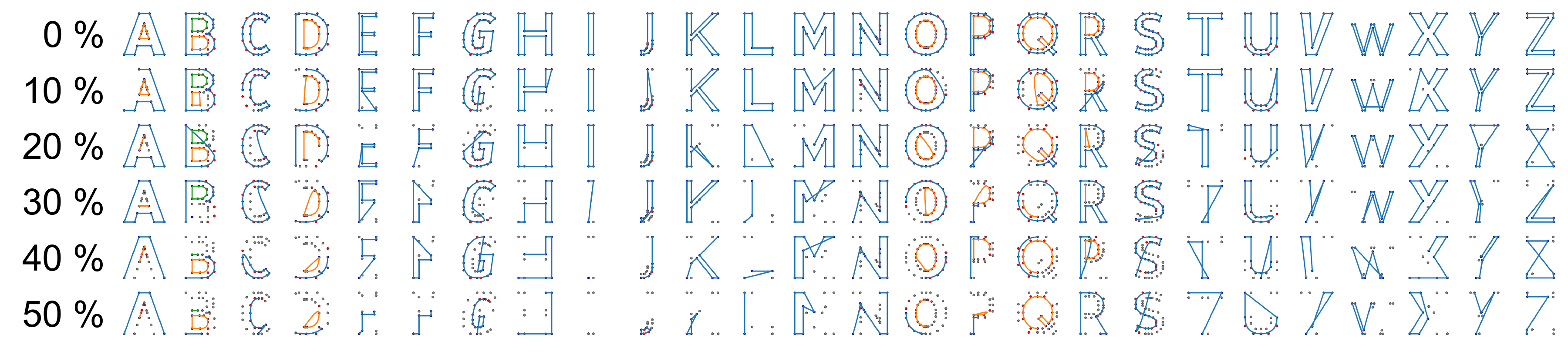}\vspace{-1mm}
(a)~{Input (Burst deletion)}
\vspace{4mm}

\includegraphics[width=1.0\linewidth]{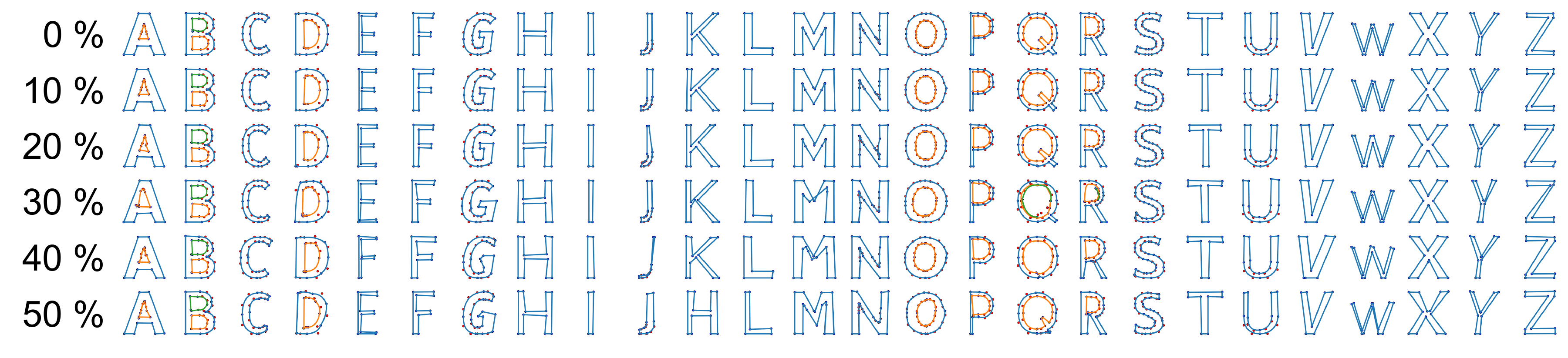}\vspace{-1mm}
(b)~{Proposed method}
\vspace{4mm}

\includegraphics[width=1.0\linewidth]{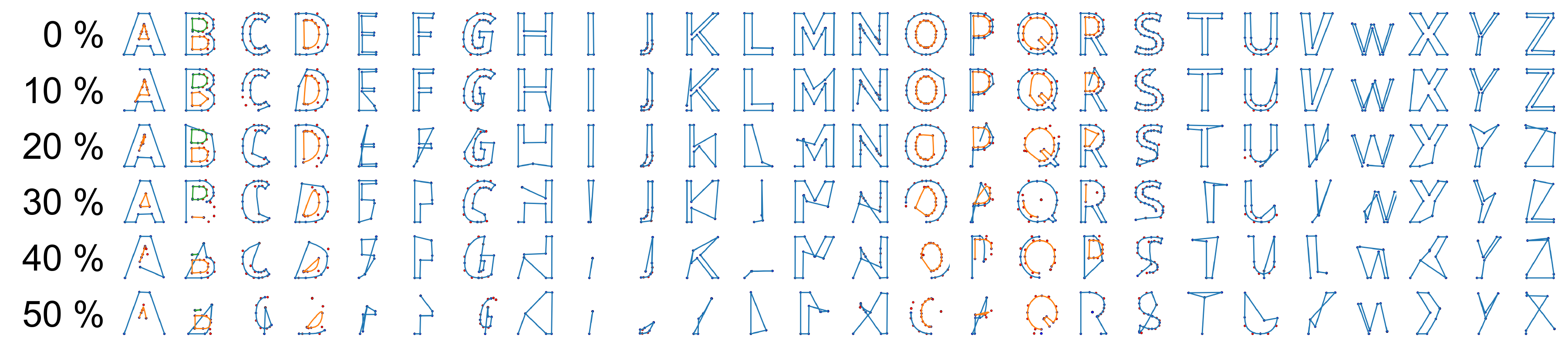}\vspace{-1mm}
(c)~{Comparative method}

\caption{The results from burst deletion. The columns are characters in the same font style, and the rows are the same missing percentages. The first row is the ground truth complete contour and second to sixth row have a deletion ratio of 10\% to 50\%, respectively. The colors are the different contours of each character.}
\label{fig:burst_results}
\end{figure*}

\subsection{Qualitative Evaluation}

Fig.~\ref{fig:random_results} shows the generated results for random deletion and Fig.~\ref{fig:burst_results} shows the generated results for the burst deletion.
In both cases, the proposed method outputs contour data closer to the correct contour data.
Notably, the proposed method was able to reconstruct many of the characters, despite having only 50\% of the original control points. 

Supporting Fig.~\ref{fig:gragh}, the figures show that the proposed method was more effective on the data with burst deletion than random deletion. 
It was able to complete the contours more accurately. 
On the random deletion characters in Fig.~\ref{fig:random_results}~(b), there is more overlap between the contours. 
For example, the ``B'' and ``O'' at high deletion rates had more overlapping orange and blue contours.
Furthermore, ``G,'' ``C,'' ``N,'' ``S,'' and ``J'' have more regions where the contours are too thin, when compared to Fig.~\ref{fig:burst_results}~(b).
This indicates that the random deletion removes more important information that cannot be recovered by the proposed method when compared to burst deletion. 
The proposed method is able to reconstruct lost segments from burst deletion more accurately based on  the intact features.

\begin{figure}[!t]
\centering
\includegraphics[width=1.0\linewidth]{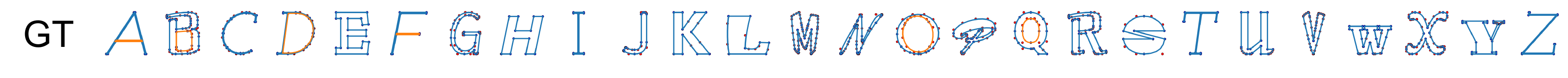}\vspace{-1mm}
(a)~{Ground truth}
\vspace{4mm}

\includegraphics[width=1.0\linewidth]{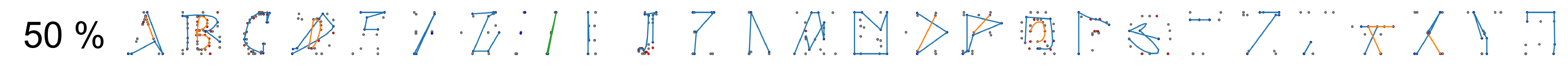}\vspace{-1mm}
(b)~{Input (Random deletion), 50\% }
\vspace{4mm}

\includegraphics[width=1.0\linewidth]{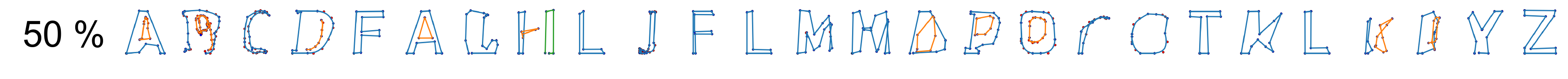}\vspace{-1mm}
(c)~{Proposed method (Random deletion)}
\vspace{4mm}

\includegraphics[width=1.0\linewidth]{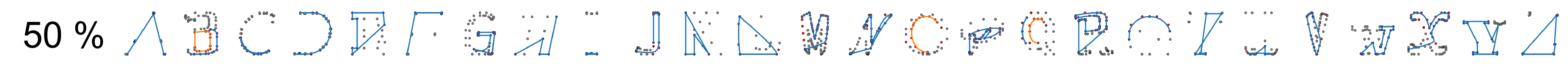}\vspace{-1mm}
(d)~{Input (Burst deletion), 50\% }
\vspace{4mm}

\includegraphics[width=1.0\linewidth]{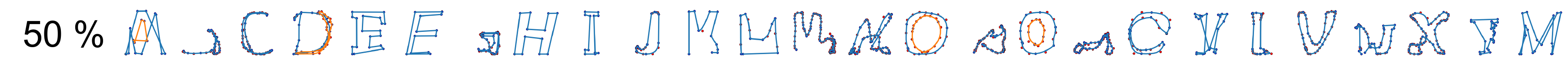}\vspace{-1mm}
(e)~{Proposed method (Burst deletion)}

\caption{Examples of Poor Reconstructions by the Proposed Method}
\label{fig:failure_case}
\end{figure}

Examples of poor reconstructions are shown in Fig.~\ref{fig:failure_case}. 
In these examples, the proposed method was unable to reconstruct the characters accurately. 
Many of the burst deletion characters only had small regions to base the reconstructions on. 
Thus, the proposed method was unable to extrapolate the characters.

\begin{figure}[!t]

\centering
\includegraphics[width=1.0\linewidth]{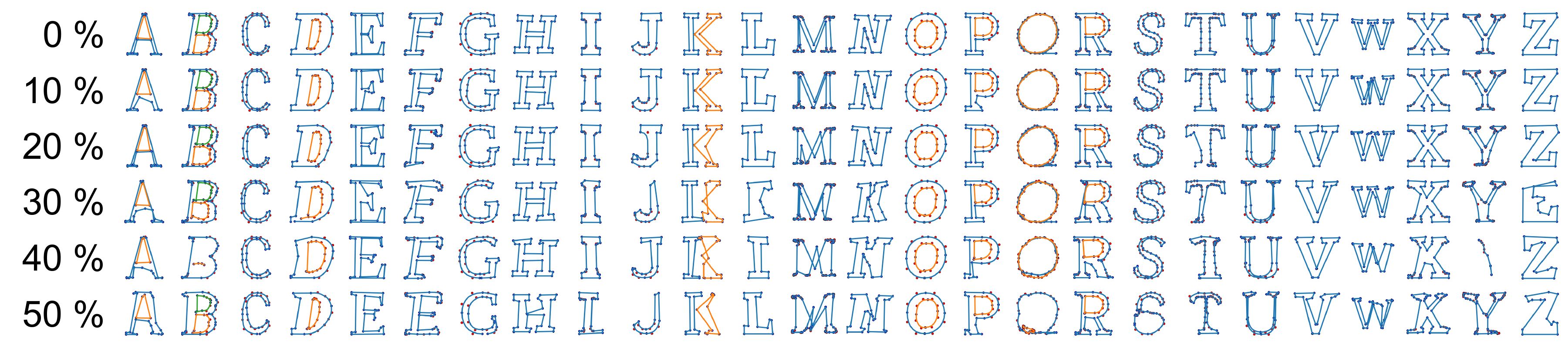}\vspace{-1mm}
\label{fig:burst_SERIF_predicted}
(a)~{Serif}
\vspace{4mm}

\centering
\includegraphics[width=1.0\linewidth]{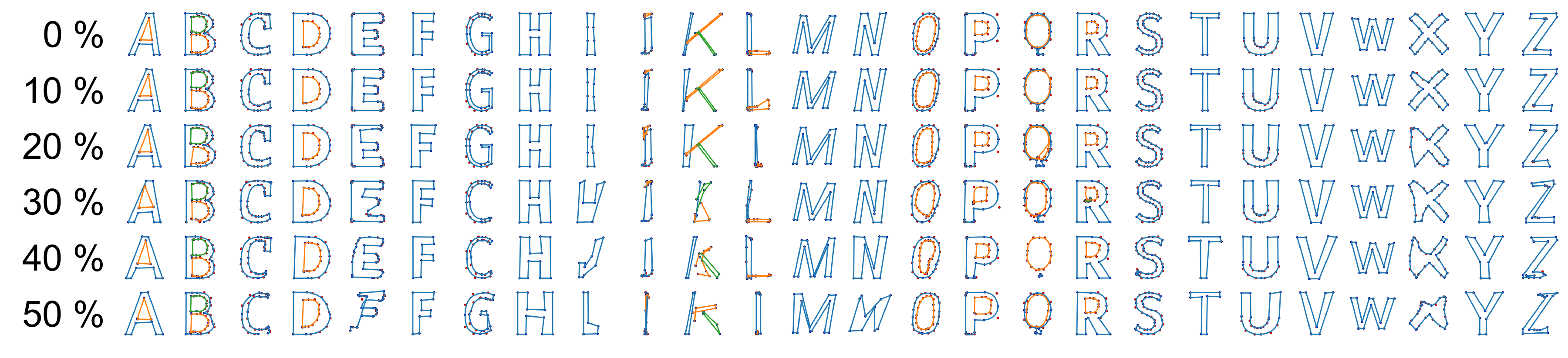}\vspace{-1mm}
\label{fig:burst_SANS_SERIF_predicted}
(b)~{Sans-Serif}
\vspace{4mm}

\centering
\includegraphics[width=1.0\linewidth]{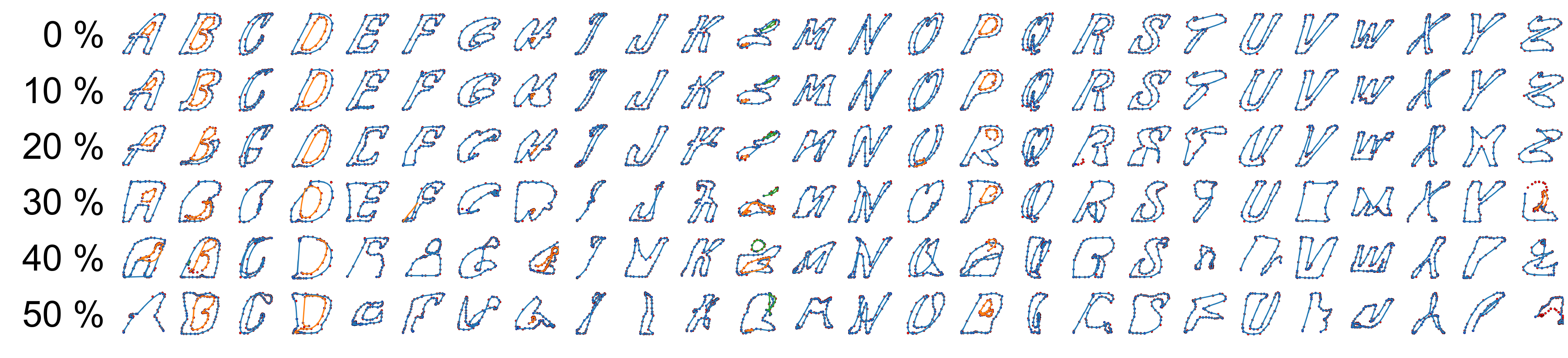}\vspace{-1mm}
\label{fig:burst_HANDWRITING_predicted}
(c)~{Handwriting}
\vspace{4mm}

\centering
\includegraphics[width=1.0\linewidth]{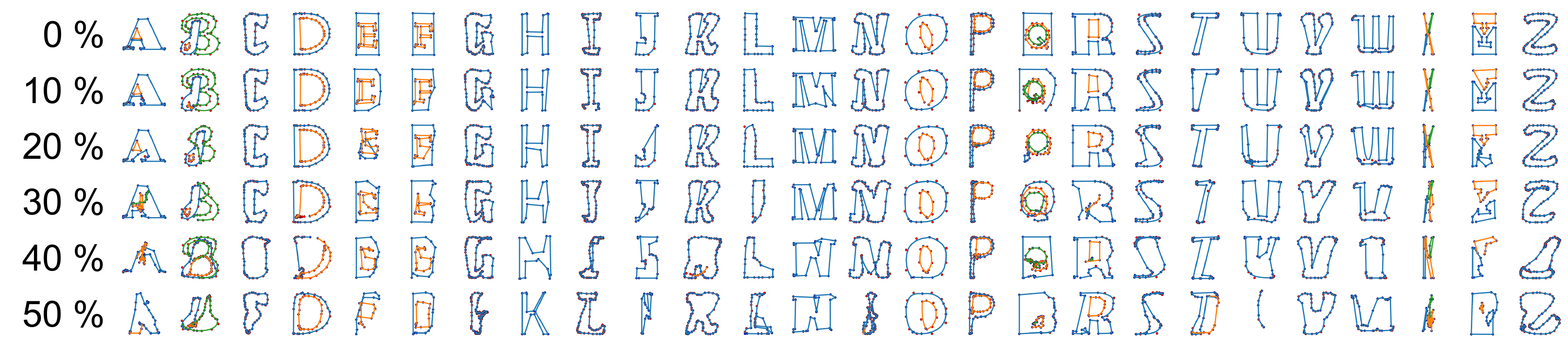}\vspace{-1mm}
\label{fig:burst_DISPLAY_predicted}
(d)~{Display}

\caption{Examples of generated results using the proposed method on burst deletion separated by font category.}
\label{fig:burst_predicted}
\end{figure}

\subsection{Comparison Between Styles}
The degree of difficulty varies depending on the font category. 
For example, Fig.~\ref{fig:burst_predicted} has examples of reconstructed contours from the proposed method when burst deletion is used. 
Each row is different deletion rates with 10\% at the top and 50\% on the bottom of each subsection.
As seen in Fig.~\ref{fig:burst_predicted} (a) and (b), reconstruction was relatively easy for Serif and Sans-Serif, despite having very little information to reconstruct the character from.  
This is due to Serif and Sans-Serif fonts having less variation compared to Display and Handwriting fonts. 
Display fonts in particular have a large variation between fonts, thus the results of 50\% deletion rate are poor.

\subsection{Comparison Between Characters}\label{sec:compare_char}

\begin{figure}[t]
\begin{minipage}[b]{0.48\textwidth}
    \centering
    \includegraphics[width=1.0\linewidth]{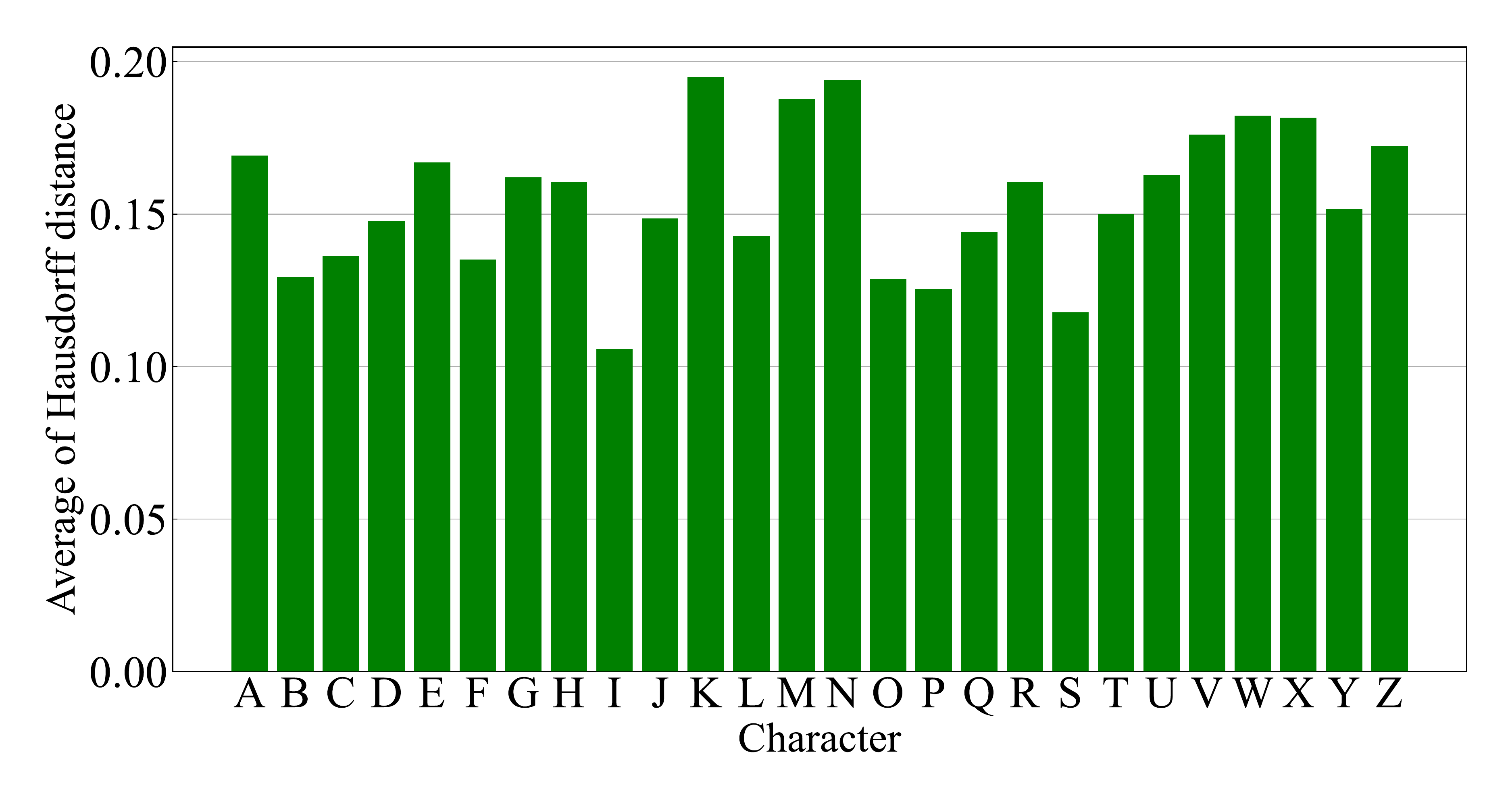}\\
    (a) Hausdorff distance with random deletion
\end{minipage}
 \hspace{0.05\linewidth}
\begin{minipage}[b]{0.48\textwidth}
    \centering
    \includegraphics[width=1.0\linewidth]{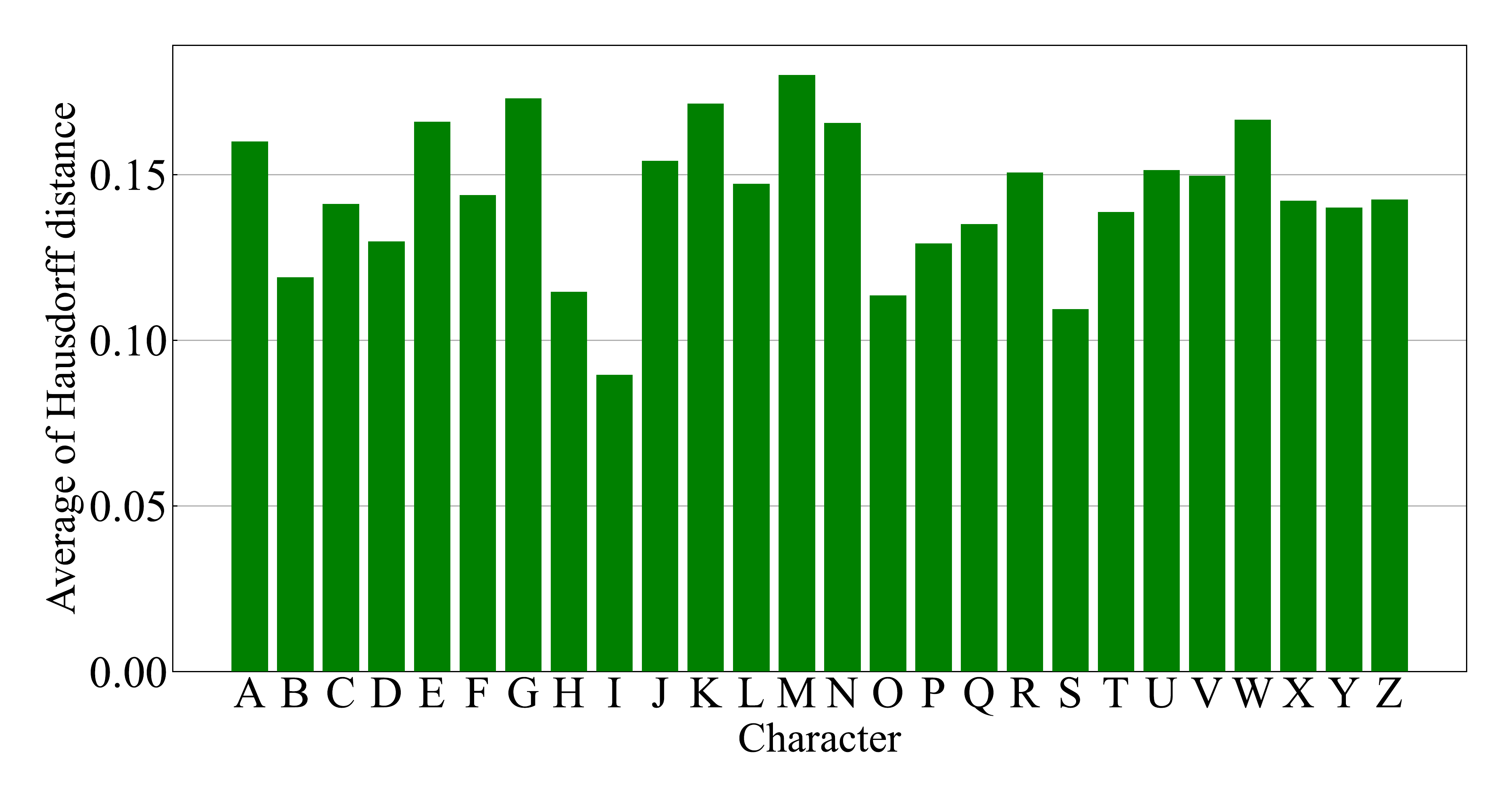}\\
    (b) Hausdorff distance with burst deletion
\end{minipage}
\caption{Average distance between the proposed method and the ground truth when separated by character class}
\label{fig:compare_char}
\end{figure}

Certain characters tend to be more difficult to reconstruct. 
A comparison of characters is shown in Fig.~\ref{fig:compare_char}. 
In the figure, the Hausdorff distance is calculated for each character with inputs with random deletion and burst deletion. 
The average from 10-50\% was used for the evaluation. It can be seen that there is a difference in accuracy for each character. The characters that obtained high accuracy with the proposed model were ``I,'' ``B,'' ``P,'' and ``S.'' On the contrary, ``K,'' ``U,'' ``V,'' and ``W' were the characters that did not obtain high accuracy.
In addition, ``X'' was one of the characters for which the difference between the random deletion and the burst deletion was significant.

\section{Conclusion}

In this paper, we proposed the use of a Transformer-based model to tackle contour completion for vector characters. 
The proposed method is indication-free in that it does not require any indication of where, if, or how much the contour is missing. 
It simultaneously solves missing point identification and value estimation.
Specifically, we proposed an Encoder-Decoder Transformer with a multitask output and loss.

To evaluate the proposed method, we constructed two datasets of characters in a variety of fonts. 
The characters are represented by sequences of control points. 
Then, in order to test the contour completion ability of the proposed method, we perform corruptions of the vector characters. 
To do this, in one dataset, we randomly delete points and in the other, we remove a continuous segment from the character. 
In this paper, we demonstrated that the proposed method can effectively reconstruct the corrupted characters. 
Future work could include the introduction of an image-based loss function to improve the quality of the output contour data.

\newpage

%
%
\bibliographystyle{splncs04}
\bibliography{myrefs}

\end{document}